\documentclass[twocolumn,aps,superscriptaddress,showpacs,floatfix]{revtex4}
\usepackage{graphicx}
\usepackage{booktabs}
\usepackage{amssymb,bm,mathrsfs,bbm,amscd}
\usepackage[tbtags]{amsmath}
\usepackage{lastpage}
\usepackage{CJK}\usepackage{epsfig}

\begin{document}
\begin{CJK*}{GBK}{song}

%===================================================================================================================

\title{Decomposition of EOS of Asymmetric Nuclear Matter into Different Spin-isospin Channels
}

\author{Wei Zuo\footnote{zuowei@impcas.ac.cn}}
\affiliation{Institute of Modern Physics, Chinese Academy of
Sciences, Lanzhou 730000, China}
 \author{Shan-Gui ZHOU}\affiliation{Institute of Theoretical Physics, Chinese Academy of sciences,
Beijing, China}
\author{ Jun-Qing LI}\affiliation{Institute of Modern Physics, Chinese Academy of
Sciences, Lanzhou 730000, China}
 \author{ En-Guang ZHAO}\affiliation{Institute of Theoretical Physics, Chinese Academy of sciences,
Beijing, China}
 \author{ Werner Scheid}\affiliation{Institute f\"ur Theoretische Physik der
Justus-Liebig-Universit\"at, D-35392 Giessen, Germany}

\begin{abstract}
We investigate the equation of state of asymmetric nuclear matter
and its isospin dependence in various spin-isospin $ST$ channels
within the framework of the Brueckner-Hartree-Fock approach extended
to include a microscopic three-body force (TBF). It is shown that
the potential energy per nucleon in the isospin-singlet $T=0$
channel is mainly determined by the contribution from the tensor
$SD$ coupled channel. At high densities, the TBF effect on the
isospin-triplet $T=1$ channel contribution turns out to be much
larger than that on the $T=0$ channel contribution. At low densities
around and below the normal nuclear matter density, the isospin
dependence is found to come essentially from the isospin-singlet
$SD$ channel and the isospin-triplet $T=1$ component is almost
independent of isospin-asymmetry. As the density increases, the
$T=1$ channel contribution becomes sensitive to the
isospin-asymmetry and at high enough densities its
isospin-dependence may even become more pronounced than that of the
$T=0$ contribution. The present results may provide some microscopic
constraints for improving effective nucleon-nucleon interactions in
nuclear medium and for constructing new functionals of effective
nucleon-nucleon interaction based on microscopic many-body theories.

{\bf keyword:}
equation of state, asymmetric nuclear matter,
 decomposition into spin-isospin channels, three-body force,
 Brueckner-Hartree-Fock approach

\end{abstract}
%\footnotetext[0]{\hspace*{-2em}\small\centerline{\thepage\ --- \pageref{LastPage}}}%

%\begin{multicols}{2}

 \maketitle
%---------------------------------------------------------------

%%%%%%%%%%%%%%%%%%%%%%%%%%%%%%%%%%%%
\section{Introduction}
%%%%%%%%%%%%%%%%%%%%%%%%%%%%%%%%%%%%%

 Effective nucleon-nucleon (NN)
interactions such as the Skyrme and Skyrme-like interactions play an
important role in predicting the properties of finite
nuclei~\cite{vautherin:1972,friedrich:1986,dobaczewski:1996,%
goriely:2002,goriely:2003,lesinski:2007,brito:2007}, nuclear matter
and neutron
stars~\cite{onsi:2002,stone:2003,stone:2006,meissner:2007},
nucleus-nucleus interaction potential~\cite{denisov:2002,wang:2006}
and fission barriers~\cite{goriely:2007}. The parameters of the
effective interactions are usually constrained by the ground state
properties of stable nuclei and the saturation properties of nuclear
matter, and thus they are shown to be quite successful for
describing nuclear phenomena related to nuclear system not far from
the normal nuclear matter density ($\rho_0=0.17$fm$^{-3}$) at small
isospin-asymmetries. However, as soon as the density deviates from
the normal nuclear matter density and the isospin-asymmetry becomes
large, the discrepancy among the predictions of the
Skyrme-Hartree-Fock (SHF) approach by adopting different Skyrme
parameters could be extremely large~\cite{babrown:2000,chen:2005}.
As for the isospin dependence of single-particle properties,
different Skyrme parameters may lead to an opposite isospin
splitting of the neutron and proton effective masses in neutron-rich
nuclear matter even at densities around
$\rho_0$~\cite{lesinski:2006}. In order to improve the predictive
power of the Skyrme interaction at high densities and large isospin
asymmetries, some work was done in recent years to constrain the
Skyrme parameters by fitting the bulk properties of asymmetric
nuclear matter obtained by the SHF approach to those predicted by
the microscopic many-body theories. For example, in
Ref.~\cite{chabanat} Chabanat {\it et al.} proposed a number of sets
of Skyrme parameters by reproducing the equation of states (EOSs) of
symmetric nuclear matter and pure neutron matter predicted by the
microscopic variational approach~\cite{pudliner:1995}. In
Ref.~\cite{cao:2006}, the authors constructed the LNS parameters for
the Skyrme interaction by fitting to the EOS of asymmetric nuclear
matter and the neutron/proton effective mass splitting in
neutron-rich matter around saturation density obtained within the
Brueckner-Hartree-Fock
 (BHF) approach extended to include a microscopic three-body force
(TBF)~\cite{zuo:2002,zuo:2005}. Although these recent
parametrizations of Skyrme interaction can reproduce fairly well the
EOSs of symmetric nuclear matter and pure neutron matter predicted
by microscopic approaches(variational method and BHF approach), the
deviation from the microscopic results is shown to become
significantly large even for symmetric nuclear matter as soon as the
EOS is decomposed into different spin-isospin
channels~\cite{lesinski:2006}. Therefore it is of interest to
investigate the EOS of asymmetric nuclear matter and its isospin
dependence in various spin-isospin $ST$ channels within the
framework of the microscopic BHF approach for a deeper understanding
of the mechanism of the isospin dependence of the nuclear EOS and
for providing more rigorous and elaborate microscopic constraints
for effective NN interactions.

In the present paper, we shall decompose the EOS of asymmetric
nuclear matter into various spin-isospin $ST$ channels by using the
BHF approach extended to include a microscopic three-body force. We
shall discuss particularly the isovector part and the isospin
dependence of the EOS of asymmetric nuclear matter in different
spin-isospin $ST$ channels. The obtained results are expected to
provide some useful information for constraining the spin-isospin
properties of effective NN interactions.

\section{Theoretical approaches}

Our present calculation is based on the Brueckner
theory~\cite{day:1967}. The Brueckner approach for asymmetric
nuclear matter and its extension to include a microscopic TBF can be
found in Ref.~\cite{zuo:2002,zuo:1999}. Here we simply give a brief
review for completeness. The starting point of the BHF approach is
the reaction $G$-matrix, which satisfies the following isospin
dependent Bethe-Goldstone (BG) equation,
\begin{eqnarray}
G(\rho, \beta, \omega )&=& \upsilon_{NN} +\upsilon_{NN}
 \nonumber \\ &\times&
\sum_{k_{1}k_{2}}\frac{ |k_{1}k_{2}\rangle Q(k_{1},k_{2})\langle
k_{1}k_{2}|}{\omega -\epsilon (k_{1})-\epsilon (k_{2})}G(\rho,
\beta, \omega ) ,
\end{eqnarray}
where $k_i\equiv(\vec k_i,\sigma_1,\tau_i)$, denotes the momentum,
the $z$-component of spin and isospin of a nucleon, respectively.
$\upsilon_{NN}$ is the realistic NN interaction, $\omega$ is the
starting energy. The asymmetry parameter is defined as
$\beta=(\rho_n-\rho_p)/\rho$, where $\rho, \rho_n$, and $\rho_p$
denote the total, neutron and proton number densities, respectively.
In solving the BG equation for the $G$-matrix, the continuous
choice~\cite{jeukenne:1976} for the auxiliary potential $U(k)$ is
adopted since it provides a much faster convergence of the hole-line
expansion than the gap choice~\cite{song:1998}. Under the continuous
choice, the auxiliary potential describes the BHF mean field felt by
a nucleon during its propagation in nuclear
medium~\cite{lejeune:1978}.

The BG equation has been solved in the total angular momentum
representation~\cite{zuo:1999}. By using the standard
angular-averaging scheme for the Pauli operator and the energy
denominator, the BG equation can be decoupled into different partial
wave $\alpha=\{JST\}$ channels~\cite{baldo:1999}, where $J$ denotes
the total angular momentum, $S$ the total spin and $T$ the total
isospin of a two-particle state.

For the NN interaction, we adopt the Argonne $V_{18}$ ($AV_{18}$)
two-body interaction~\cite{wiringa:1995} plus a microscopic based on
the meson-exchange current approach~\cite{grange:1989}. The
parameters of the TBF model have been self-consistently determined
so as to reproduce the $AV_{18}$ two-body force by using the
one-boson-exchange potential model~\cite{zuo:2002}. The TBF contains
the contributions from different intermediate virtual processes such
as virtual nucleon-antinucleon pair excitations, and nucleon
resonances ( for details, see Ref.~\cite{grange:1989}). The TBF
effects on the EOS of nuclear matter and its connection to the
relativistic effects in the DBHF approach have been reported in
Ref.~\cite{zuo:2002}.

The TBF contribution has been included by reducing the TBF to an
equivalently effective two-body interaction via a suitable average
with respect to the third-nucleon degrees of freedom according to
the standard scheme~\cite{grange:1989}. The effective two-body
interaction ${\tilde v}$ can be expressed in $r$-space
as\cite{zuo:2002}
\begin{equation}
\begin{array}{lll}
 \langle\vec r_1 \vec r_2| {\tilde v} |
\vec r_1^{\ \prime} \vec r_2^{\ \prime} \rangle = \displaystyle
\frac{1}{4} {\rm Tr}\sum_n \int {\rm d} {\vec r_3} {\rm d} {\vec
r_3^{\ \prime}}\phi^*_n(\vec r_3^{\ \prime})(1-\eta(r_{23}'))
 \\[5mm]\displaystyle
 \times
 \displaystyle (1-\eta(r_{13}' ))W_3(\vec r_1^{\ \prime}\vec r_2^{\
\prime} \vec r_3^{\ \prime}|\vec r_1 \vec r_2 \vec r_3)
 (1-\eta(r_{13}))\\[3mm] \times (1-\eta(r_{23})) \phi_n(r_3)
\end{array}\label{eq:TBF}
\end{equation}
where the trace is taken with respect to the spin and isospin of the
third nucleon. The function $\eta(r)$ is the defect function. Since
the defect function is directly determined by the solution of the BG
equation\cite{grange:1989}, it must be calculated self-consistently
with the $G$ matrix and the s.p. potential $U(k)$\cite{zuo:2002} at
each density and isospin asymmetry. It is evident from
Eq.(\ref{eq:TBF}) that the effective force ${\tilde v}$ rising from
the TBF in nuclear medium is density dependent. A detailed
description and justification of the method can be found in
Ref.~\cite{grange:1989}.

\section{Results and Discussion}

\begin{center}%\vspace{-2mm}
\begin{figure}\includegraphics[width=8.5cm]{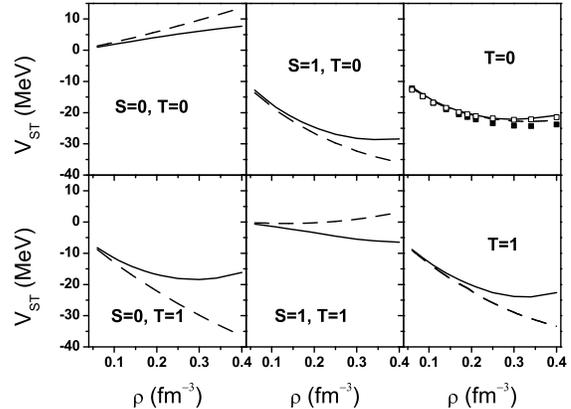}\\
\caption{
 Decomposition of the potential part of the EOS (i.e., potential
 energy per nucleon vs. density) of
 symmetric nuclear matter into various spin-isospin $ST$ channels.
 Dashed curves: results obtained by adopting purely the $AV_{18}$
two-body NN interaction; Solid curves: results obtained by adopting
the $AV_{18}$ plus the TBF; Full and empty squares: potential energy
per nucleon in the tensor $^3SD_2$ channel calculated in the two
cases of including and not including the TBF, respectively.
 }
%\label{f:b00Vst}
\end{figure}\end{center}
In Fig.~1 we display the potential part of the EOS of symmetric
nuclear matter (i.e., the potential energy per nucleon as a function
of density) in various spin-isospin channels of $ST =00, 10, 01,
11$, and $T=0, 1$. The solid curves are obtained by including the
TBF, while the dashed curves are the results by adopting purely the
$AV_{18}$ two-body interaction. The empty and filled squares
indicate the contributions of the $T=0\ $ $^3SD_2$ tensor channel in
the two cases by adopting the pure $AV_{18}$ two-body interaction
and the $AV_{18}$ plus the TBF. We notice from the figure that the
absolute values of the potential energy per nucleon in the two even
partial wave channels ($ST=10$ and $ST=01$) are much lager than
those in the two odd channels ($ST=00$ and $ST=11$). It is also seen
that the potential energy in the isospin-singlet $T=0$ channel is
mainly determined by the contribution from the tensor $^3SD_2$
channel. At relatively low densities up to about $0.25$fm$^{-3}$,
the attraction of the potential energy in both $T=0$ and $T=1$
channels increases monotonically as a function of density. While, in
the high density region of $0.25\le\rho\le 0.4$fm$^{-3}$, the
density dependence of the potential energy per nucleon in the $T=0$
channel becomes very weak (slowly increases with density) in both
cases of including and not including the TBF.

The TBF effect can be seen clearly by comparing the solid curves
with the corresponding dashed curves. At small densities, the TBF
effect turns out to be reasonably weak. As density increases, the
TBF effect becomes significant. The TBF contributions are shown to
be repulsive for the two even partial wave channels ($ST=10$ and
$ST=01$), while they are attractive for the two odd channels
($ST=00$ and $ST=11$). When the TBF is not included, the two
components in the $ST=10$ and $ST=01$ channels are strongly
attractive and their attraction increases monotonically with
density. The TBF makes these two components much less attractive at
high densities and as a consequence these two components become
slowly increasing functions of density at high enough densities. In
the case of not including the TBF, the components in the two odd
partial wave channels are repulsive. Inclusion of the TBF makes the
potential energy per nucleon in the spin-triplet odd channel
($ST=11$) become negative from positive. When the TBF is included,
the potential energies per nucleon in both the isospin singlet $T=0$
and isospin triplet $T=1$ channels are shown to be almost
independent of density (solid curves in the right two panels of
Fig.~1) in the high-density region of $0.25\le\rho\le 0.4$fm$^{-3}$
and thus the kinetic part, which increases monotonically with
density as $\rho^{2/3}$, of the energy per nucleon is expected to
determine the stiffness of the EOS at high enough densities
($\rho\ge0.25$fm$^{-3}$).

It is worth mentioning that the two spin-triplet $ST=10$ and $ST=11$
components obtained in the present paper are quite different from
those predicted by Baldo {\it et al.} (Crosses in Fig.6 of
Ref.~\cite{lesinski:2007}). The two components predicted by Baldo
{\it et al.} are slowly increasing functions of density, while the
$ST=11$ component obtained in the present paper decreases with
density slowly. The discrepancy between the two calculations is
mainly caused by the two different TBFs adopted. In the calculation
of Baldo {\it et al.} the phenomenological Urbana
TBF~\cite{baldo:1997} was adopted.

\begin{center}%\vspace{-2mm}
\begin{figure}\includegraphics[width=8.5cm]{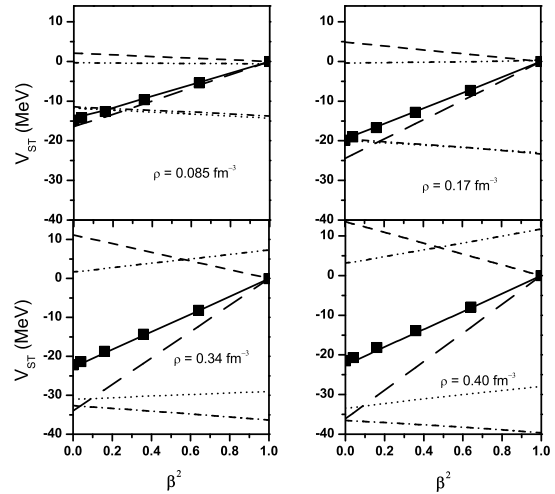}\\
\caption{
 Isospin-asymmetry dependence of the contributions to the EOS of
 asymmetric nuclear matter from various spin-isospin channels for
 several values of density $\rho=0.085, 0.17, 0.34$ and 0.4fm$^{-3}$,
 respectively. Solid curves: $T=0$ channel; long-dashed curves: $ST=00$ channel;
 short-dashed curve: $ST=10$ channel; dotted curves: $T=1$ channel;
 dot-dashed curves: $ST=01$ channel; double-dot-dashed curves:
 $ST=11$ channel; squares: $SD$ tensor channel. The results are obtained in the case of not
 including the TBF.
 }
%\label{f:EOS2bf}
\end{figure}\end{center}

Now we turn to discuss the isospin dependence of the various
spin-isospin $ST$ channel contributions to the EOS of asymmetric
nuclear matter. In Fig.~2 we show the potential energy per nucleon
in the $ST=00, 10, 01, 11$ and $T=0, 1$ channels vs. $\beta^2$ for
several values of density. The results of Fig.~2 have been obtained
by adopting purely the $AV_{18}$ two-body force. It is seen from the
figure that the isospin dependence of the isospin-triplet
($ST=01,11$, and $T=1$) components is extremely weak as compared
with the isospin-singlet ($ST=00,10$, and $T=0$) components, and
thus the isospin dependence of the potential part of the EOS of
asymmetric nuclear matter is determined essentially by the
contribution of the isospin-singlet $T=0$ channel. As for the $T=0$
channel, the isospin dependence turns out to stem almost completely
from the contribution of the tensor $SD$ coupled channel (comparing
the solid curves and the corresponding squares), while the
contributions of the other isospin-singlet channels cancel almost
completely.

In order to discuss the TBF effects, we give in Fig.~3 the
$\beta^2$-dependence of the potential energy per nucleon in various
spin-isospin channels obtained by including the TBF. It is seen that
at relatively low densities around and below the normal nuclear
matter density, the TBF effect is reasonably small. The components
of the potential energy in the isospin-triplet channels ($ST=01,
11$, and $T=1$) are almost independent of $\beta$ and thus the
isospin dependence of the potential energy is essentially determined
by that of its component in the isospin $T=0$ channel. As the
density increases, the isospin dependence of the components in the
isospin-triplet channels becomes significant since the TBF effect
increases continuously as the density increases. We may notice that
the TBF affects the isospin-dependence of the isospin-triplet
channels much more significantly as compared with the
isospin-singlet channels. At high densities, the TBF effect on the
isospin-singlet $T=0$ channel is rather weak, but it affects
strongly the isospin-dependence of the potential energy per nucleon
in the isospin-triplet $T=1$ channel. At high enough densities (for
example, $\rho=0.4$fm$^{-3}$) the sensitivity of the $T=1$ channel
potential energy may even become more pronounced than that in the
$T=0$ channel (comparing the solid and dotted curves in the
lower-right panel of Fig.~3) due to the TBF effect. By comparing
Fig.~3 with Fig.~2, we can see that the isospin dependence of the
$T=1$ channel potential energy at high densities comes mainly from
the TBF contribution.

\begin{center}%\vspace{-2mm}
\begin{figure}\includegraphics[width=8.5cm]{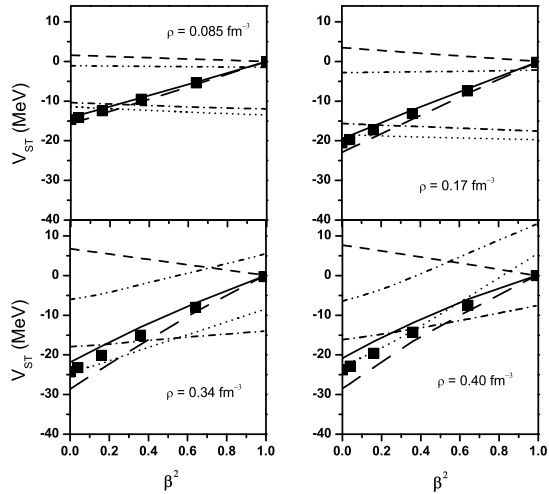}\\
\caption{
 The same as in Fig.~2 but the results are obtained by
 including the TBF.
 }
%\label{f:EOS3bf}
\end{figure}\end{center}

%%%%%%%%%%%%%%%%%%%%%%%%%%%%%%%%%%%%%
\section{Summary}
%%%%%%%%%%%%%%%%%%%%%%%%%%%%%%%%%%%%
In the present paper, we have investigated the EOS of asymmetric
nuclear matter by decomposing its potential part into various
spin-isospin $ST$ channels by using the BHF approach extended to
include a microscopic three-body force. We have de-composed the
isovector part of the EOS of asymmetric nuclear matter into
different $ST$ channels and discussed particularly the isospin
dependence of the EOS of asymmetric nuclear matter in different
spin-isospin $ST$ channels. The potential energy per nucleon in the
isospin-singlet $T=0$ channel and its isospin-dependence are shown
to be determined to a large extent by the contribution of the tensor
$SD$ coupled channel. At low densities around and below the normal
nuclear matter density, the TBF effect is fairly weak and the
potential energy in the isospin-triplet $T=1$ channel is almost
independent of isospin-asymmetry. Consequently, the isospin
dependence is found to come essentially from the isospin-singlet
$SD$ channel. At high densities, the TBF effect on the
isospin-triplet $T=1$ channel contribution turns out to be much
larger than that on the $T=0$ channel contribution. As the density
increases, the $T=1$ channel contribution becomes more and more
sensitive to the isospin-asymmetry and at high enough densities its
isospin-dependence may even become more pronounced than that of the
$T=0$ contribution.

The present results may shed light for understanding the origin of
the isospin dependence of the EOS of asymmetric nuclear matter and
provide some useful information for constraining the effective NN
interactions and their isospin dependence in asymmetric nuclear
medium.
\\
%\section*{Acknowledgments}
\section*{Acknowledgments}
{Supported by the National Natural Science
Foundation of China
(10875151, 10575119, 10435010, 10575036, 10705014, 10811130077), the
Knowledge Innovation Project(KJCX3-SYW-N2) of Chinese Academy of
Sciences, the Major State Basic Research Developing Program of China
under No. 2007CB815004, CAS/SAFEA International Partnership Program
for Creative Research Teams (CXTD-J2005-1), and the DFG of Germany.
}

%\end{multicols}
%%%%%%%%%%%%%%%%%%%%%%%%%%%%%%%%%%%%%%%%%%%

\end{CJK*}
\end{document}